\documentclass[prd,aps,a4paper,superscriptaddress,twocolumn,nofootinbib]{revtex4}
\usepackage{graphicx}
\usepackage{color}
\usepackage{dcolumn}
\usepackage{bm}
\usepackage{slashed}
\usepackage{amsmath}
\usepackage{latexsym}
\usepackage{amssymb}
\usepackage{mathrsfs}
\usepackage{amsfonts}
\usepackage{url}
\allowdisplaybreaks
\begin{document}
\title{Validating the Effective-One-Body Numerical-Relativity Waveform Models
for Spin-aligned Binary Black Holes along Eccentric Orbits}

\author{Xiaolin Liu}
\affiliation{Department of Astronomy, Beijing Normal University,
Beijing 100875, China}
\author{Zhoujian Cao
\footnote{corresponding author}} \email[Zhoujian Cao: ]{zjcao@amt.ac.cn}
\affiliation{Department of Astronomy, Beijing Normal University,
Beijing 100875, China}
\author{Lijing Shao} 
\affiliation{Kavli Institute for Astronomy and Astrophysics, Peking University, Beijing 100871, China}

\begin{abstract}
Effective-one-body (EOB) numerical-relativity (NR) waveform models for spin-aligned binary black holes (BBHs), known as the SEOBNR waveform models, are based on the EOB theoretical framework and NR simulations. SEOBNR models have played an important role in the LIGO scientific collaboration (LSC) gravitational wave (GW) data analysis for both signal search and parameter estimation. SEOBNR models for quasi-circular orbits have evolved through version 1 to version 4 by extending their validity domain and including
more NR results. Along another direction, we recently extended SEOBNRv1 model to SEOBNRE model which is valid for spin-aligned BBH coalescence along eccentric orbits. In this paper we validate this theoretical waveform model by comparing them against the numerical relativity simulation bank, Simulating eXtreme Spacetimes (SXS) catalog. In total, 278 NR waveforms are investigated which include binaries with large eccentricity; large spin and large mass ratio. Our SEOBNRE can model the NR waveforms quite well. The fitting factor for most of the 278 waveforms is larger than 99\%. It indicates that the SEOBNRE model could be used as template waveforms for eccentric spin-aligned BBH coalescence.
\end{abstract}

\maketitle

\section{Introduction}
During the O1 and O2 observations, LIGO/Virgo have detected 11 gravitational wave (GW) events \cite{2018arXiv181112907T}. Besides them, additional possible ones are reported by some external groups \cite{Nitz_2019a,Nitz_2019b,Magee_2019,PhysRevD.100.023007,PhysRevD.100.023011}. Matched filtering technique has played a very important role in all of these detections. In order to make it work, an accurate gravitational waveform model is needed. The effective-one-body numerical-relativity (EOBNR) model works very well for LIGO/Virgo detections. Although most analysis has assumed that the compact binary coalescence in the LIGO/Virgo frequency band admits a vanishingly small eccentricity, it is still interesting to ask about the actuall eccentricity for these events. Recently we have extended the EOBNR model to describe eccentric compact binary coalescence, and it is named the SEOBNRE model \cite{PhysRevD.96.044028}. The authors in Ref.~\cite{Romero-Shaw:2019itr} have used the SEOBNRE model to estimate the eccentricity of the LIGO/Virgo GW events.

Buonanno and Damour proposed the original idea of effective one body (EOB) method for binary black hole (BBH) in general relativity \cite{PhysRevD.59.084006}. Later Buonanno, Pan and others for the first time \cite{PhysRevD.76.104049} combined the EOB method with numerical-relativity (NR) to obtain the EOBNR model families for BBH coalescence. Aiming for a faithful waveform template for LIGO/Virgo, SEOBNRv1 \cite{PhysRevD.86.024011}, SEOBNRv2 \cite{PhysRevD.89.061502}, SEOBNRv3 \cite{PhysRevD.95.024010} and SEOBNRv4 \cite{PhysRevD.95.044028} were consequently constructed by extending the validity domain and including more NR simulations. Recently EOBNR models have also been developed to describe the waveform of binary neutron stars \cite{PhysRevD.95.104036,PhysRevD.98.104052,PhysRevD.99.044051,PhysRevD.99.044007}. Even for gravitational wave memory, EOBNR model is also available \cite{Fav09b,Cao16}.

In Ref.~\cite{PhysRevD.96.044028} we extended the SEOBNRv1 model to the SEOBNRE model which can describe eccentric BBH coalescence. Besides our SEOBNRE model, other groups also developed different theoretical models for eccentric BBH systems \cite{PhysRevD.97.024031,PhysRevD.98.044015,PhysRevD.95.024038,PhysRevD.93.124061,PhysRevD.93.064031,PhysRevD.91.084040}. The major difference between our SEOBNRE model and these models is that our SEOBNRE model does not take the adiabatic approximation. Like other EOBNR models, our SEOBNRE model can produce full waveforms including the inspiral, the merger and the ringdown.

The authors in Ref.~\cite{PhysRevD.96.104048} also used EOB method to construct waveform models for binary systems. There are two different aspects between our SEOBNRE model and their model. Firstly they did not combine the NR results to improve their model. Secondly they used adiabatic approximation and took eccentricity directly as a state variable to describe the binary system.

The SEOBNRv1 model behaves quite bad if the spin of the black hole is large. The SEOBNRv4 cures this limit \cite{PhysRevD.95.044028}. Regarding to eccentric BBHs, our SEOBNRE model extends the limit of quasi-circular systems \cite{PhysRevD.96.044028}. Since our SEOBNRE model is developed based on SEOBNRv1, we doubt it may also admit limitation on high spins. In the current paper we will investigate this problem and find out the limitation. This finding will provide clues for us to improve our SEOBNRE model in the future.

Throughout this paper we will use the unit system where $c=G=1$. We denote the masses of the two component black holes $m_1$ and $m_2$ respectively, and assume $m_1\geq m_2$. We denote the total mass $M=m_1+m_2$, the mass ratio $q\equiv m_1/m_2$ and the symmetric mass ratio $\eta=m_1m_2/M^2$. We use $\vec{S}_1$ and $\vec{S}_2$ to denote the spin of the two black holes. Then we have dimensionless spin parameters $\vec{\chi}_1=\vec{S}_1/m_1^2$ and $\vec{\chi}_2=\vec{S}_2/m_2^2$. Furthermore we assume the orbital angular momentum of the binary points to the $z$ direction at the initial time. Then we denote denote $\chi_{1z}=S_{1z}/m_1^2$ and $\chi_{2z}=S_{2z}/m_2^2$. We notate the effective spin $\chi_{\rm eff}=(m_1\chi_{1z}+m_2\chi_{2z})/M=(q\chi_{1z}+\chi_{2z})/(1+q)$ \cite{PhysRevLett.113.151101}, and the anti-symmetric spin $\chi_{A}=(m_1\chi_{1z}-m_2\chi_{2z})/M=(q\chi_{1z}-\chi_{2z})/(1+q)$ \cite{Cao16}.

This paper is arranged as following. We describe the comparison setup between theoretical model and NR waveform in the next section. After that the comparison results between the NR waveform and the generated waveforms by SEOBNRE model are presented in the Sec.~\ref{secIV}. We find that SEOBNRE fits the NR waveforms better than 99\% when the orbit eccentricity at frequency $Mf_0=0.002$ is less than 0.55 and the spin ``hang-up" effect is not too strong. This finding not only provides us the limitation of the SEOBNRE model but also validate the SEOBNRE waveform model for mildly eccentric BBHs with mild spins. Based on this confidence, we use our SEOBNRE model to calibrate the referenced eccentricity of the NR waveforms in the Sec.~\ref{secV}. Finally we give a discussion and a summary in the last section.

\section{Comparison setup}\label{secII}
In the current paper we consider only $(2,2)$ spin weighted spherical harmonic mode. Suppose we have two waveforms needed to be compared $h_1(t)$ and $h_2(t)$. Since the waveforms considered in the current paper all include inspiral, merger and ringdown, there is a maximal value for the amplitude of each waveform respect to time, the so-called amplitude peak. Firstly we align the time of the two waveforms to let their maximal amplitudes appear at $t=0$. Then assume the waveform $h_1$ starts from time $t_{11}<0$ and ends at time $t_{21}>0$. For waveform $h_2$ these two times are $t_{12}<0$ and $t_{22}>0$ respectively. Then we take $t_1=\max(t_{11},t_{12})$ and $t_2=\min(t_{21},t_{22})$. Based on the times $t_1$ and $t_2$ we cut parts of the two waveforms beyond the time range $(t_1,t_2)$. Following this procedure we get two equal time duration waveforms. And we compare these two remaining equal-length waveforms.

Now we assume we have two equal-time-duration waveforms $h_1(t)$ and $h_2(t)$. We define inner product of them as,
\begin{align}
\langle h_1|h_2\rangle&=4\max_{t_0,\phi_0}\Re\left[\int_{f_{min}}^{f_{max}}\tilde{h}_1\tilde{h}_2^*e^{i(2\pi ft_0+\phi_0)}df\right],\label{equation1}
\end{align}
where the ``$\tilde{}$" means the Fourier transformation, the ``${}^*$" means taking the complex conjugate, and ``$\Re$" means taking the real part; $t_0$ and $\phi_0$ are the initial time and initial phase used to match the two waveforms. Many previous works, like Refs.~\cite{PhysRevD.95.044028,PhysRevD.96.044028,PhysRevD.98.104052}, used LIGO's sensitivity curve to define the inner product. Differently, here we concern more about the theoretical model behavior itself, so we essentially use a uniform sensitivity in (\ref{equation1}). The motivation for this choice is aiming to make EOBNR models work not only for LIGO but also for future space-based detectors, such as LISA \cite{bender1998lisa,amaro2012low,audley2017laser}, Taiji \cite{gong2011scientific} and Tianqin \cite{luo2016tianqin}. The upper bound of the integration $f_{max}$ corresponds to the sampling rate in the waveforms. The lower bound of the integration $f_{min}$ corresponds to the time duration of the waveforms. The same choice was taken in our previous work \cite{PhysRevD.96.044028} when we constructed the SEOBNRE model.
\begin{figure}
\begin{tabular}{c}
\includegraphics[width=0.5\textwidth]{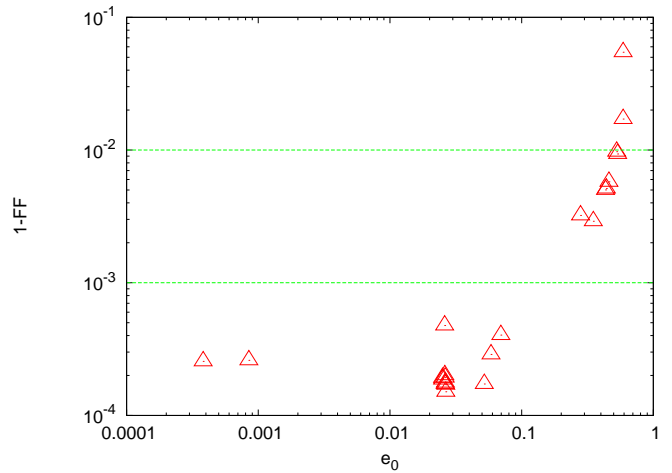}
\end{tabular}
\caption{Comparison of NR waveforms and SEOBNRE model for equal-mass nonspinning BBH coalescence, as a function of the eccentricity $e_0$. The eccentricity $e_0$ shown in the horizontal axis is obtained by fitting the SEOBNRE waveform through adjusting the eccentricity at reference frequency $Mf_0=0.002$.}\label{fig1}
\end{figure}

According to our initial alignment, $t=0$ corresponds to the amplitude peak of the waveform. However, there exist another alignment that makes the matching better. In this situation, we align again the two waveforms according to the fitting procedure in Eq.~(\ref{equation1}) by adjusting the alignment time $t_0$.

Based on the inner product (\ref{equation1}), we have the fitting factor
\begin{align}
{\rm FF}&\equiv\frac{\langle h_1|h_2\rangle}{\|h_1\|\cdot\|h_2\|},\\
\|h\|&\equiv\sqrt{\langle h|h\rangle}.
\end{align}

\begin{figure*}
\begin{tabular}{c}
\includegraphics[width=\textwidth]{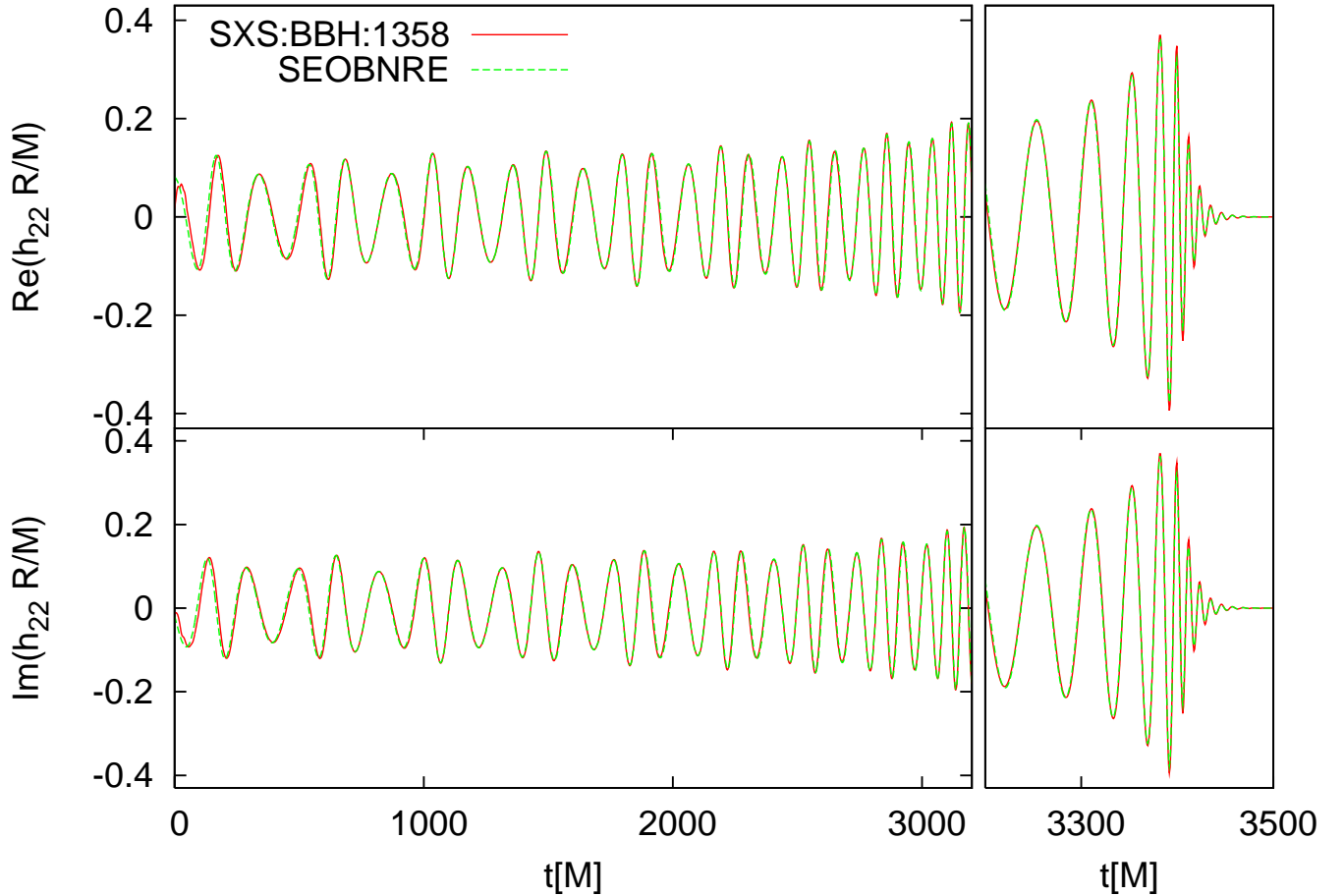}
\end{tabular}
\caption{Waveform comparison between the NR and SEOBNRE model for SXS:BBH:1358. The corresponding fitting factor is ${\rm FF}=99.4\%$. The initial eccentricity at the reference frequency $Mf_0\approx0.004$ is $e_0\approx0.22$ estimated by SXS simulation, while the fitted initial eccentricity at $Mf_0=0.002$ is $e_0=0.46$, estimated by SEOBNRE model.}\label{fig2}
\end{figure*}
\begin{figure*}
\begin{tabular}{c}
\includegraphics[width=\textwidth]{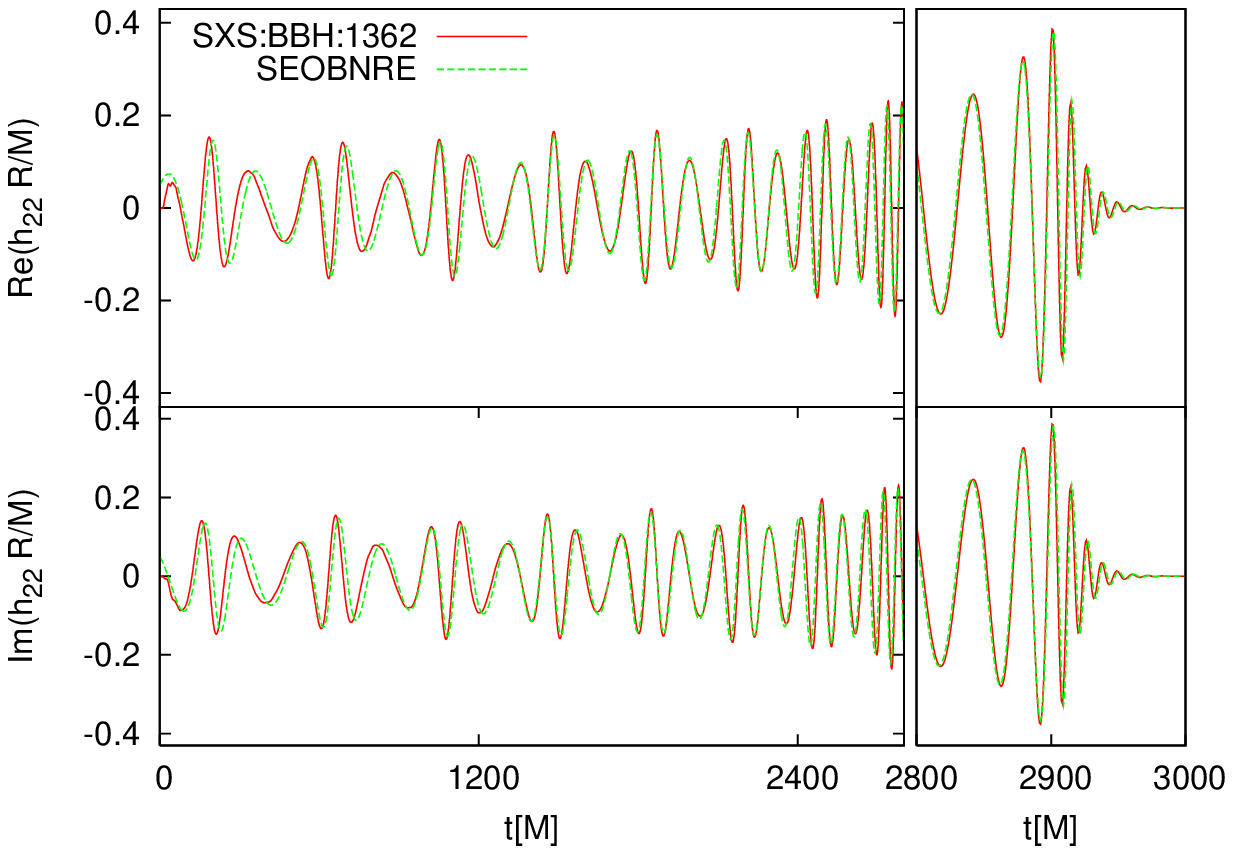}
\end{tabular}
\caption{Waveform comparison between the NR and SEOBNRE model for SXS:BBH:1362. The corresponding fitting factor is ${\rm FF}=98.3\%$. The initial eccentricity at the reference frequency $Mf_0\approx0.004$ is $e_0>1.7$ \cite{SXSBBH}, estimated by SXS simulation, while the fitted initial eccentricity at $Mf_0=0.002$ is $e_0=0.59$, estimated by SEOBNRE model.}\label{fig3}
\end{figure*}
Consider a given NR waveform $h_{22}^{NR}$, which has BBH's mass ratio $q\equiv m_1/m_2\geq1$, two individual spins $\vec{\chi}_{1,2}\equiv \vec{S}_{1,2}/m_{1,2}^2$ and a possible initial eccentricity for the orbit. For the mass ratio and black hole's spins, to make a comparison, we always adopt the values of NR waveform for the theoretical models. However, we do not use the eccentricity provided by the NR to our SEOBNRE models. This is because that, NR simulations usually start from a relatively high frequency where the eccentricity can not be well defined. Moreover, in many NR simulations, like SXS:BBH:1362, SXS:BBH:1363, SXS:BBH:1369 and others \cite{SXSBBH}, just to name a few, the eccentricity when the simulation starts can not be determined at all. Alternatively, we determine the eccentricity through,
\begin{align}
{\rm FF}&\equiv\max_{e_0}\frac{\langle h_{22}^{EOBNR}|h_{22}^{NR}\rangle}{\|h_{22}^{EOBNR}\|\cdot\|h_{22}^{NR}\|},
\end{align}
where $e_0$ is the initial eccentricity of the orbit at some given reference frequency $f_0$ of the gravitational waveform.

\section{Validating SEOBNRE models against NR waveforms}\label{secIV}

\subsection{Equal-mass nonspinning BBH cases}
For equal-mass nonspinning BBH cases, there is only one intrinsic parameter, the orbital eccentricity. As mentioned above, NR can not determine the initial eccentricity for several BBH systems. For a better comparison, we use the eccentricity obtained by fitting SEOBNRE waveforms through adjusting the eccentricity at the reference frequency $Mf_0=0.002$ to characterize the waveforms.

The resulted fitting factor is shown in Fig.~\ref{fig1}. The trend of the fitting factor as a function of the eccentricity is quite clear. This trend indicates that when the eccentricity is less than 0.2, the fitting factor is better than 99.9\%. When the eccentricity increases, the fitting factor decreases as one would expect. If the eccentricity is less than 0.55, the fitting factor is still better than 99\%. For systems with initial eccentricity $0.55<e_0<0.6$ the fitting factor will fall in the range $95\%<\text{FF}<99\%$.

We show waveform comparison examples for highly eccentric BBH systems in Fig.~\ref{fig2} for SXS:BBH:1358 with $e_0=0.46$ and in Fig.~\ref{fig3} for SXS:BBH:1362 with $e_0=0.59$.
\subsection{Nonspinning BBH cases}
\begin{figure*}
\begin{tabular}{cc}
\includegraphics[width=0.5\textwidth]{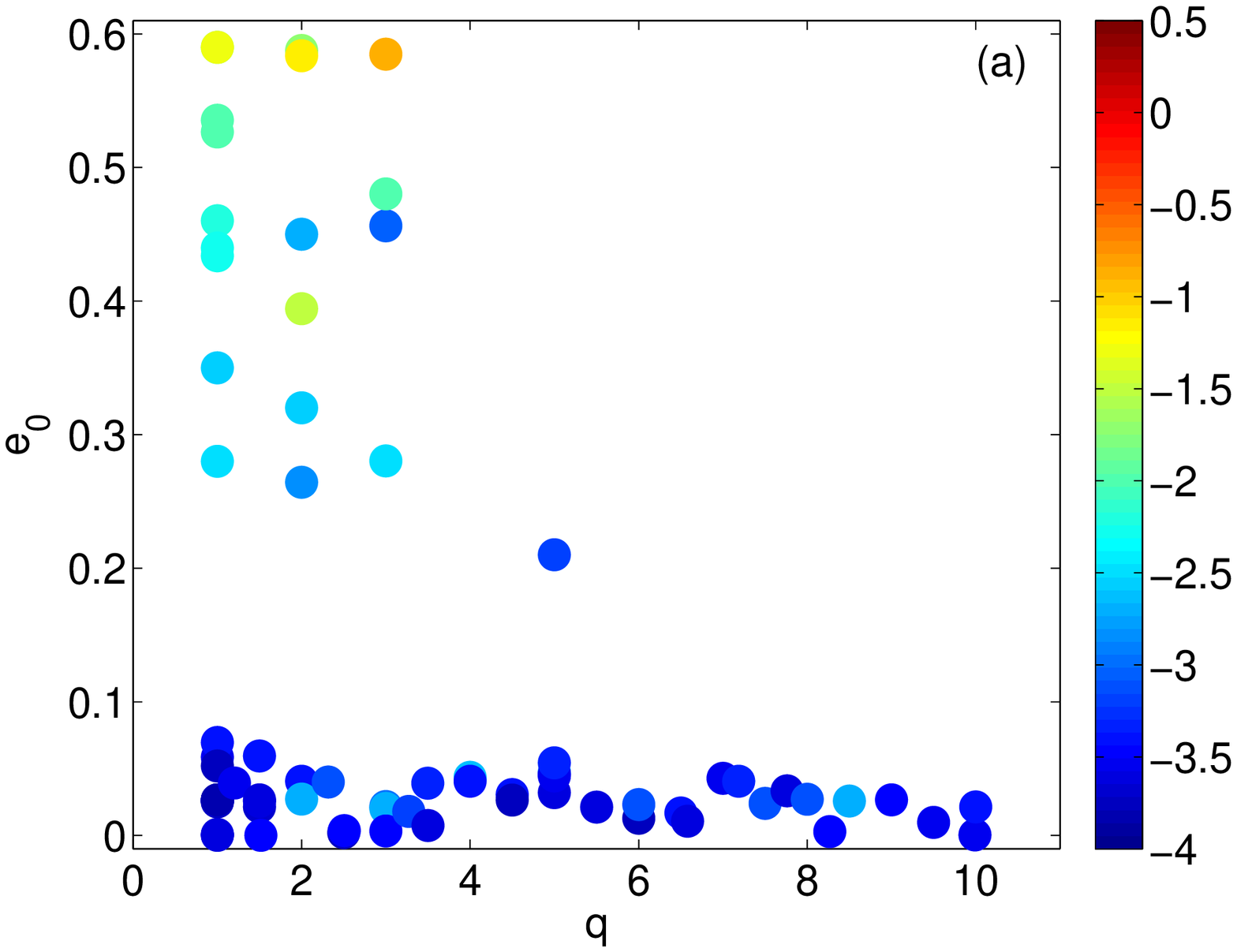}&
\includegraphics[width=0.5\textwidth]{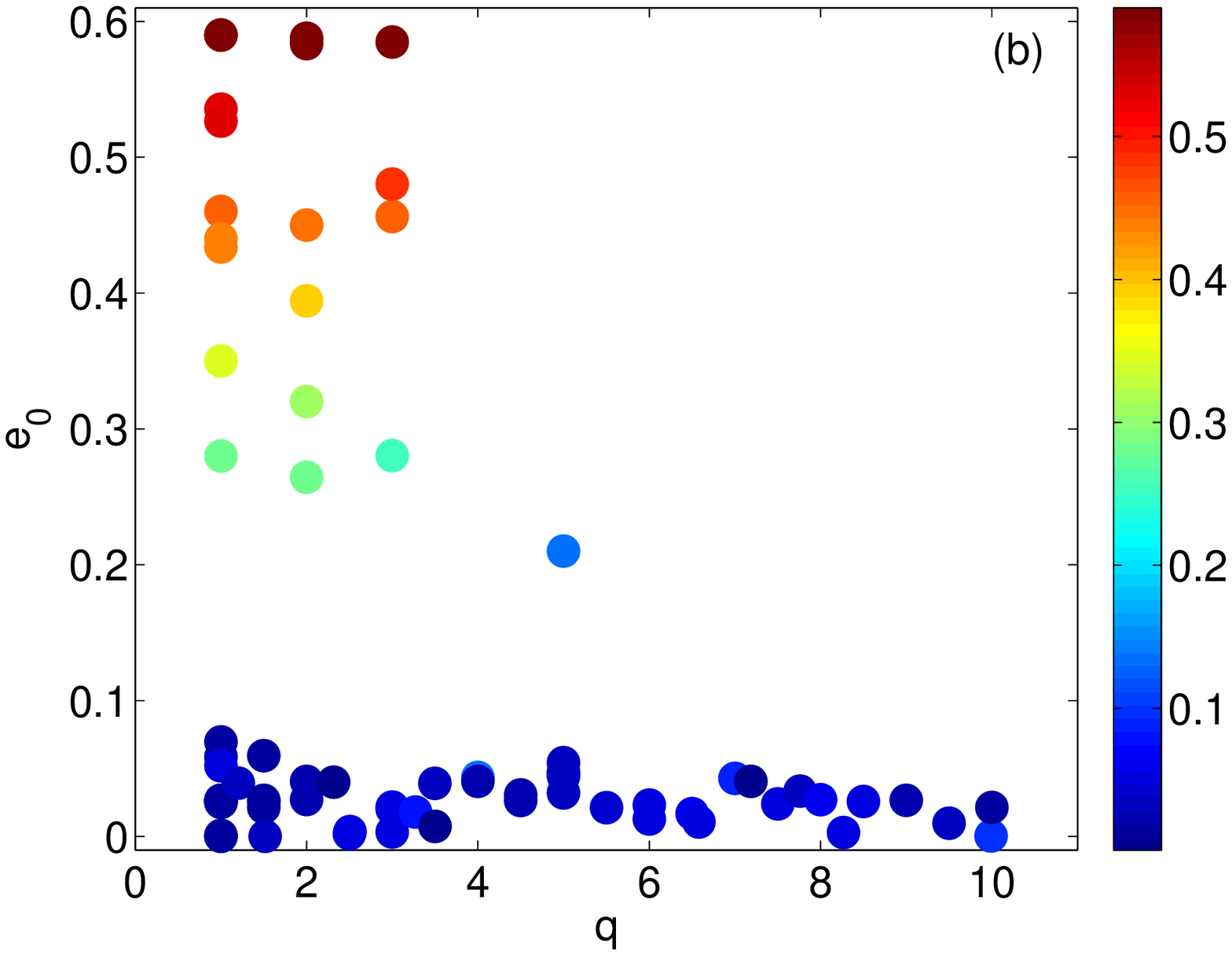}
\end{tabular}
\caption{(a) Comparison of NR waveforms and the SEOBNRE model for nonspinning BBH coalescence. The eccentricity shown in the vertical axis is obtained by fitting SEOBNRE waveforms through adjusting the eccentricity at the reference frequency $Mf_0=0.002$. The color represents $\log_{10}(1-{\rm FF})$. (b) The fraction of the energy flux contributed by the spherical harmonic modes other than $(2,\pm2)$ modes in NR, $\eta_E$, as defined in Eq.~(\ref{equation2}).}\label{fig4}
\end{figure*}

For general nonspinning BBHs, there is one more essential parameter, namely the mass ratio besides the orbital eccentricity.

The comparison results, as function of the initial eccentricity and the mass ratio are plotted in the subplot (a) of the Fig.~\ref{fig4}. For systems with $e_0<0.55$ the resulted fitting factor is better than 99\%. If the initial eccentricity becomes larger than 0.55, the fitting factor drops to 90\%. At the same time, the fitting factor is smaller when the mass ratio is larger. We suspect that this is due to the fraction of energy flux contributed by higher mode than (2,2) increases when the mass ratio and the eccentricity increase, while our current SEOBNRE model only count for (2,2) mode when we consider dissipative force. Correspondingly, in the subplot (b) of the Fig.~\ref{fig4} we plot the fraction of the energy flux contributed by the spherical harmonic modes other than $(2,\pm2)$ modes in NR waveform,
\begin{align}
\eta_E=1-\frac{E_{(2,-2)}+E_{(2,2)}}{\sum_{l=2}^8\sum_{m=-l}^{l}E_{(l,m)}}.\label{equation2}
\end{align}
\subsection{Equal-mass spin-aligned BBH cases}
\begin{figure}
\begin{tabular}{c}
\includegraphics[width=0.5\textwidth]{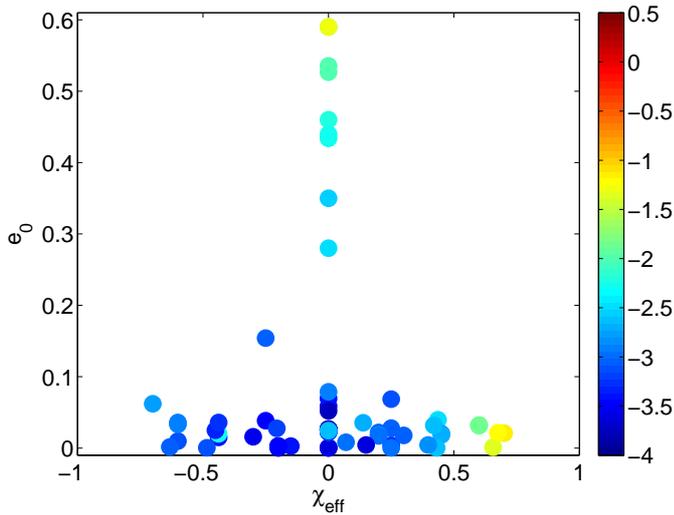}
\end{tabular}
\caption{Comparison of NR waveforms and the SEOBNRE model for equal-mass, spin-aligned BBH coalescence. The eccentricity shown in the vertical axis is obtained by fitting SEOBNRE waveforms through adjusting the eccentricity at the reference frequency $Mf_0=0.002$. The color represents $\log_{10}(1-{\rm FF})$.}\label{fig5}
\end{figure}

\begin{figure*}
\begin{tabular}{c}
\includegraphics[width=\textwidth]{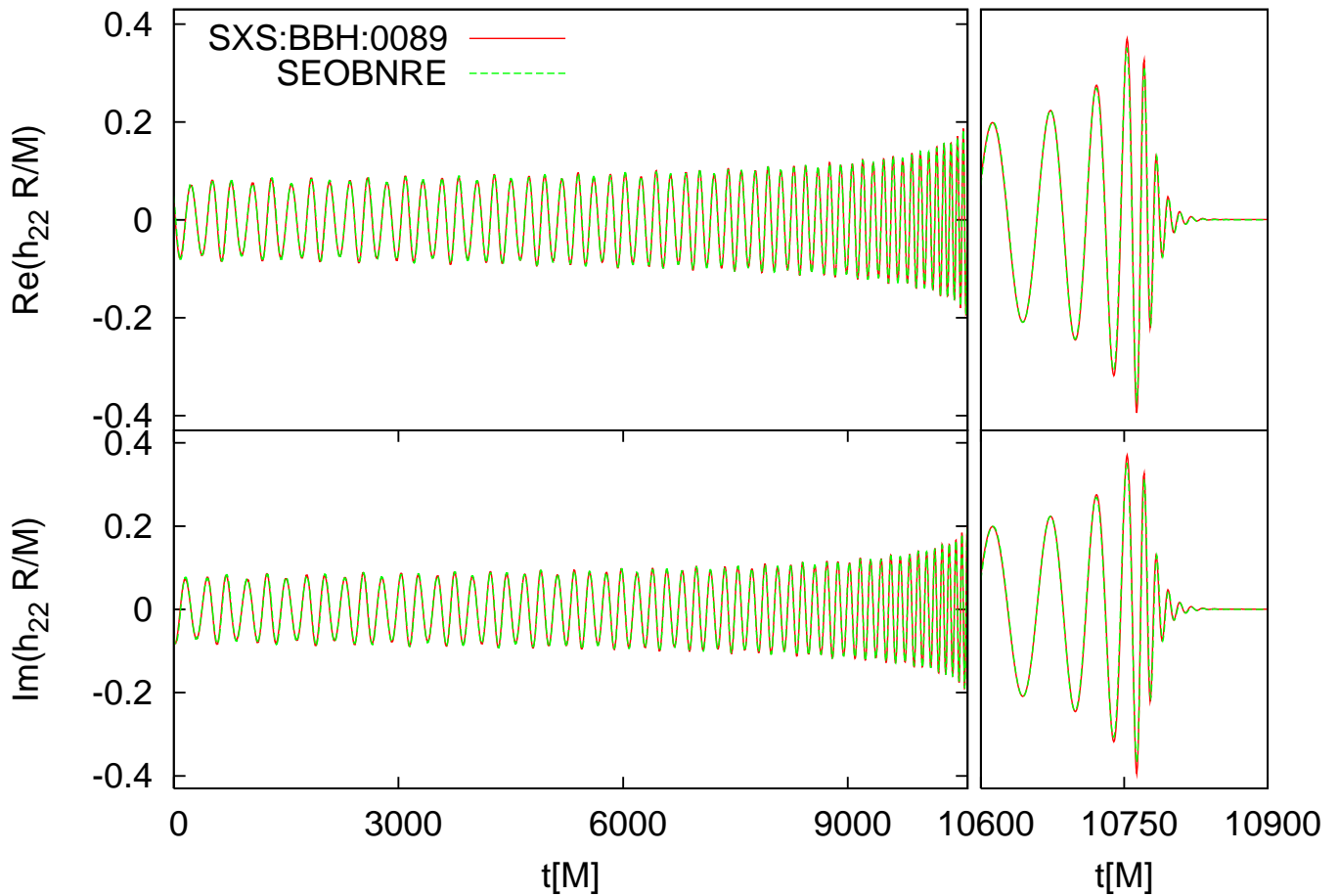}
\end{tabular}
\caption{Waveform comparison between NR and the SEOBNRE model for SXS:BBH:0089 with $\chi_{1z}=-0.5$ and $\chi_{2z}=0$. The corresponding fitting factor is ${\rm FF}=99.9\%$. The initial eccentricity at the reference frequency $Mf_0\approx0.0036$ is $e_0\approx0.06$ estimated by the SXS simulation, while the fitted initial eccentricity at $Mf_0=0.002$ is $e_0=0.154$ estimated by the SEOBNRE model.}\label{fig6}
\end{figure*}
\begin{figure*}
\begin{tabular}{c}
\includegraphics[width=\textwidth]{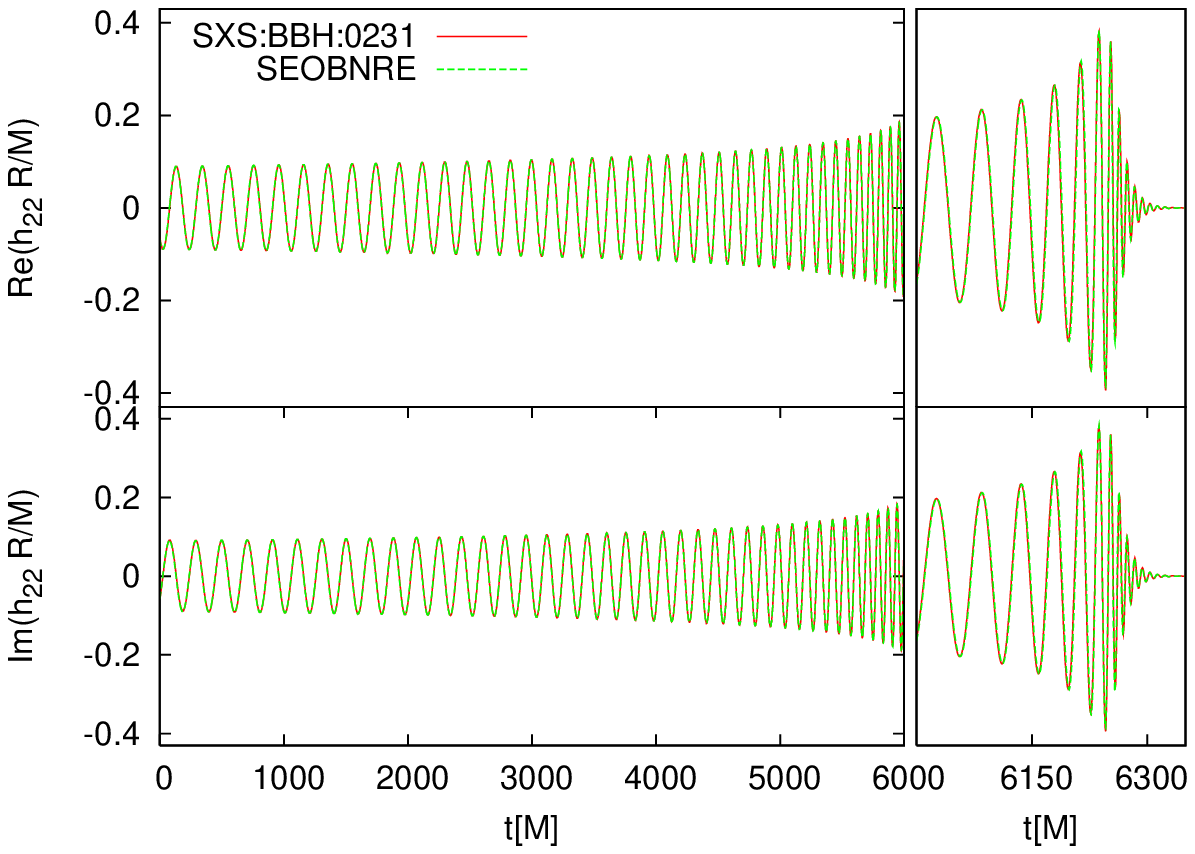}
\end{tabular}
\caption{Waveform comparison between NR and the SEOBNRE model for SXS:BBH:0321 with $\chi_{1z}=0.9$ and $\chi_{2z}=0$. The corresponding fitting factor is ${\rm FF}=99.8\%$. The initial eccentricity at the reference frequency $Mf_0\approx0.005$ is $e_0<3.4\times10^{-5}$ estimated by SXS simulation while the fitted initial eccentricity at $Mf_0=0.002$ is $e_0=0.02$ estimated by the SEOBNRE model.}\label{fig7}
\end{figure*}
For spin-aligned BBH systems, although there are three parameters involved, the effective spin $\chi_{\rm eff}$, the anti-symmetric spin $\chi_{A}$ and the orbital eccentricity, we find that the anti-symmetric spin affects little the fitting factor between NR waveforms and SEOBNRE model. Therefore, in Fig.~\ref{fig5} we plot the fitting factors between NR waveforms and the SEOBNRE model as function of the effective spin and the orbital eccentricity.

For all the investigated systems with $e_0<0.55$ and $\chi_{\rm eff}<0.52$, the resulted fitting factor is better than 99\%. We have discussed the limit of the SEOBNRE model with the initial eccentricity in the above subsections. Here we find the limitation with the spin. If the spin is anti-aligned with the orbital angular momentum, it hardly matters. But for the spin hang-up cases, when the aligned spin $\chi_{\rm eff}$ is larger than 0.52, the fitting factor will drop to about 94\%.

We plot two examples of waveform comparison for equal-mass spin-aligned BBH systems, including (i) SXS:BBH:89 with $\chi_{1z}=-0.5$, $\chi_{2z}=0$ and $\chi_{\rm eff}=-0.25$ in Fig.~\ref{fig6}, and (ii) SXS:BBH:231 with $\chi_{1z}=0.9$, $\chi_{2z}=0$ and $\chi_{\rm eff}=0.45$ in Fig.~\ref{fig7}. SXS:BBH:89 admits initial eccentricity $e_0=0.154$ at the reference frequency $Mf_0=0.002$ and fitting factor 99.9\%. For the case SXS:BBH:231 with a relatively smaller effective spin $\chi_{\rm eff}=0.45$, our SEOBNRE model can follow the spin hang-up step closely. Consequently a high fitting factor about 99.8\% is achieved.

\subsection{General spin-aligned BBH cases}
\begin{figure*}
\begin{tabular}{cc}
\includegraphics[width=0.5\textwidth]{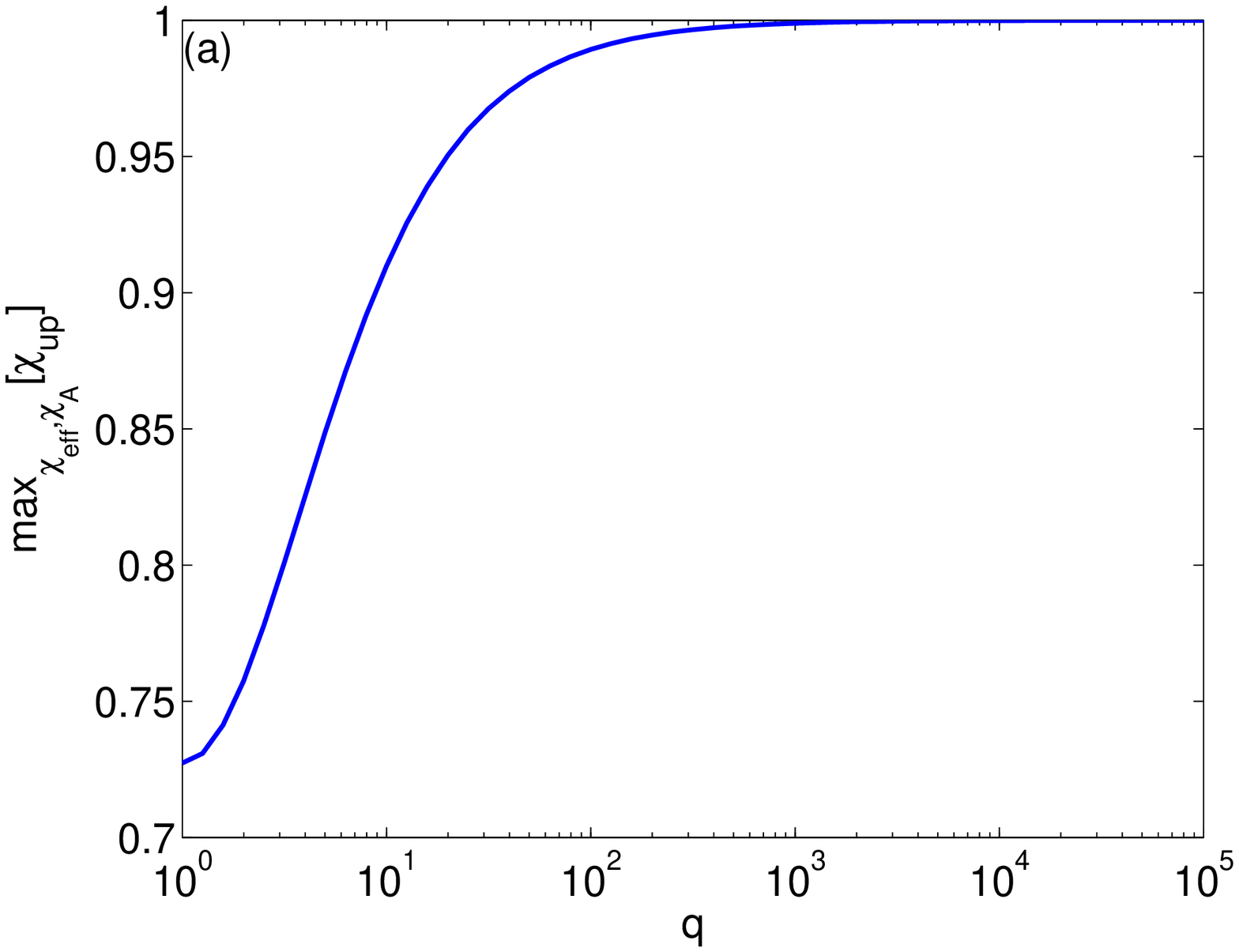}
\includegraphics[width=0.5\textwidth]{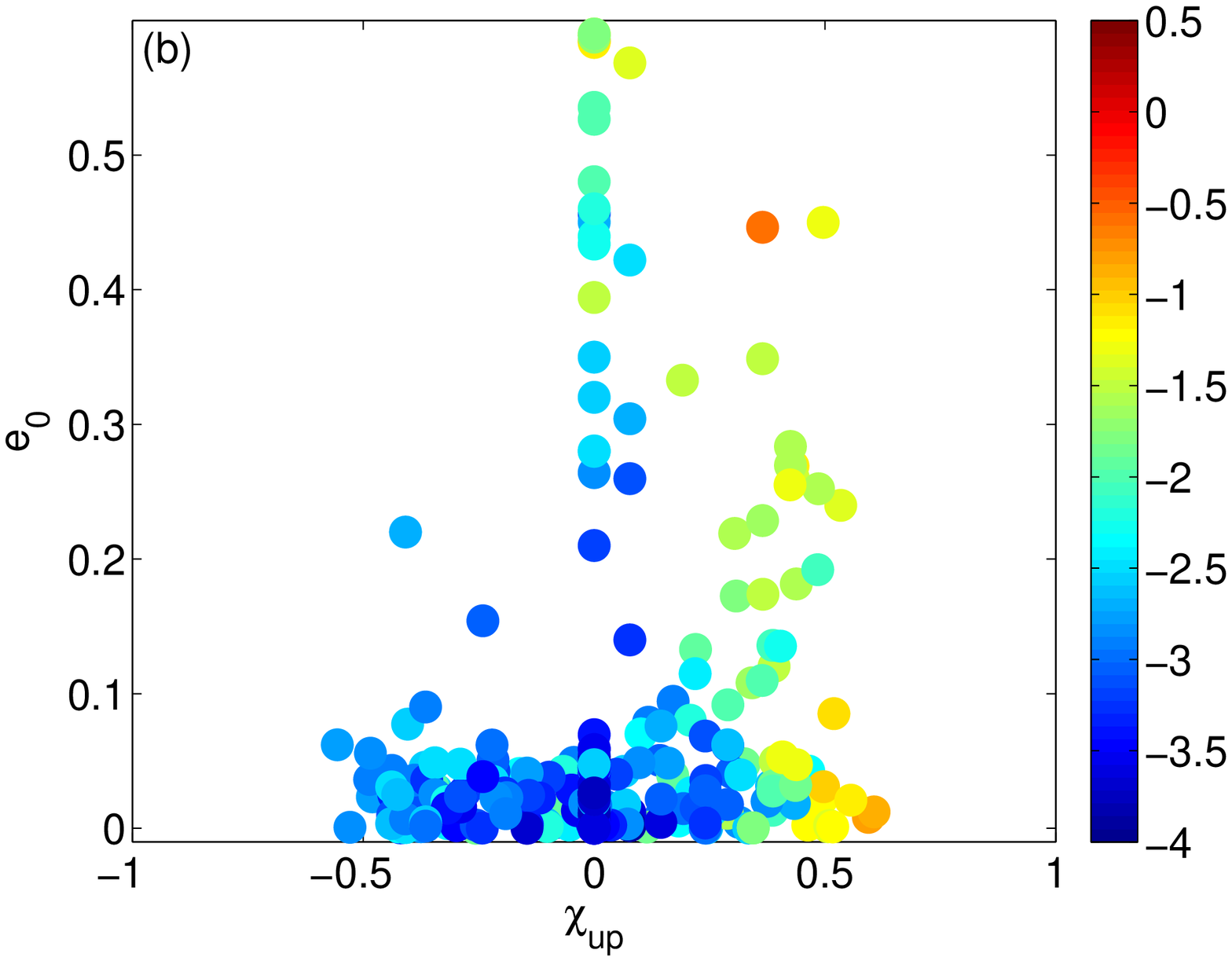}
\end{tabular}
\caption{(a) The maximal possible spin hang-up parameter $\chi_{\rm up}$--- with respect to $\chi_{\rm eff}$ and $\chi_A$--- for different mass ratio $q$. (b) Comparison of NR waveforms and the SEOBNRE model for generic BBH coalescence with any mass ratio and any aligned spins. The color represents $\log_{10}(1-{\rm FF})$. The fitting factor is shown as a function of the spin hang-up parameter $\chi_{\rm up}$ [defined in Eq.~(\ref{equation4})] and the initial eccentricity $e_0$. The eccentricity shown in the vertical axis is obtained by fitting SEOBNRE waveforms through adjusting the eccentricity at the reference frequency $Mf_0=0.002$.}\label{fig8}
\end{figure*}
At last we investigate general nonprecessing BBH systems with any mass ratio and any aligned spins. From the investigations in the above three subsections, we find that the limitation to our SEOBNRE model mainly comes from a large eccentricity and/or a strong spin hang-up effect. In order to estimate the spin hang-up effect we calculate the orbit decay rate for quasi-circular BBH systems in the Appendix. Although it is only at post-Newtonian approximation we can still get some insight from Eq.~(\ref{equation3}). It tells us the anti-symmetric spin parameter $\chi_A$ will not take effect for equal-mass systems with $\eta=0.25$. On the contrary, for unequal-mass binary systems, the anti-symmetric spin parameter $\chi_A$ will contribute to the spin hang-up effect in addition to the effective spin $\chi_{\rm eff}$. In the mean time we can see the spin hang-up contribution from the effective spin $\chi_{\rm eff}$ is independent of mass ratio.

Based on the above insight, we check the spin hang-up effect for general spin-aligned BBH systems. Equation~(\ref{equation3}) guides us to define a spin hang-up spin parameter
\begin{align}
\chi_{\rm up}\equiv \frac{8\chi_{\rm eff}+3\sqrt{1-4\eta}\chi_A}{11}.\label{equation4}
\end{align}
Later we call this parameter ``spin hang-up parameter" for short. For a given mass ratio, the maximal $\chi_{\rm up}$ is achieved when $\chi_{1z}=\chi_{2z}=1$. We plot this maximal $\chi_{\rm up}$ respect to $q$ in the subplot (a) of Fig.~\ref{fig8}.

In the subplot (b) of Fig.~\ref{fig8}, we plot the resulted fitting factor as function of $\chi_{\rm up}$ and $e_0$. As expected we find that the spin hang-up effect can not be too strong, otherwise the SEOBNRE model can not work well. Our test results indicate that, when $e_0>0.44$ and $\chi_{\rm up}>0.35$ the fitting factor will drop to about 90\%. For quasi-circular cases, when $\chi_{\rm up}$ is bigger than about 0.5, the fitting factor drops to 90\% as well.

Simply speaking, we find that if the spin hang-up effect is too strong, which can be described by the parameter $\chi_{\rm up}$, our SEOBNRE model will not work well.
\section{Reference eccentricity calibrated by the SEOBNRE model}\label{secV}
\begin{table*}
\centering
\caption{The reference eccentricity $e_0$, calibrated by SEOBNRE model, and the fitting factor $\text{FF}$ between the NR waveforms and the SEOBNRE model for spin-aligned BBH systems in the SXS simulation catalog \cite{SXSBBH}. The eccentricity $e_{0}$ is obtained by fitting the SEOBNRE model waveforms through adjusting the eccentricity at the reference frequency $Mf_0=0.002$.}
\begin{tabular}{p{1.15cm}<\centering |p{2.15cm}<\centering |p{2.15cm}<\centering || p{1.15cm}<\centering |p{2.15cm}<\centering |p{2.15cm}<\centering || p{1.15cm}<\centering |p{2.15cm}<\centering |p{2.15cm}<\centering}
\hline
SXS id & $e_{0}$ & $\text{FF}$ & SXS id & $e_{0}$ & $\text{FF}$ & SXS id & $e_{0}$ & $\text{FF}$\\
\hline
    1 &   0.02634995 &   0.99982340 &     4 &   0.03840000 &   0.99965045 &     5 &   0.00397760 &   0.99941762\\
    7 &   0.02136128 &   0.99979293 &     8 &   0.02547238 &   0.99975474 &     9 &   0.06208666 &   0.99769777\\
   12 &   0.02360445 &   0.99925434 &    13 &   0.06159936 &   0.99933327 &    14 &   0.02400026 &   0.99952261\\
   16 &   0.02400212 &   0.99962120 &    19 &   0.01095936 &   0.99883044 &    25 &   0.33280000 &   0.97100305\\
   30 &   0.02160205 &   0.99903299 &    31 &   0.22859200 &   0.97714406 &    36 &   0.00172480 &   0.99889129\\
   38 &   0.00486784 &   0.99705700 &    39 &   0.00359360 &   0.99901591 &    40 &   0.00300096 &   0.99205553\\
   41 &   0.34880000 &   0.96968799 &    45 &   0.21900160 &   0.97414722 &    46 &   0.02510400 &   0.99747785\\
   47 &   0.28356480 &   0.97443448 &    54 &   0.04723186 &   0.99733010 &    55 &   0.04548787 &   0.99964783\\
   56 &   0.03201638 &   0.99974388 &    60 &   0.00703194 &   0.99880988 &    61 &   0.05256269 &   0.95293054\\
   63 &   0.02708774 &   0.99929994 &    64 &   0.00358400 &   0.99783010 &    65 &   0.04723072 &   0.94605535\\
   66 &   0.02652813 &   0.99982970 &    67 &   0.02509041 &   0.99980812 &    68 &   0.00084800 &   0.99973975\\
   69 &   0.02603546 &   0.99952416 &    70 &   0.02622587 &   0.99982753 &    71 &   0.00003520 &   0.99973718\\
   72 &   0.02509041 &   0.99981375 &    73 &   0.00038144 &   0.99974456 &    74 &   0.05856000 &   0.99971202\\
   83 &   0.06852180 &   0.99923371 &    84 &   0.02786010 &   0.99917597 &    85 &   0.00023360 &   0.99891789\\
   86 &   0.02609894 &   0.99979937 &    87 &   0.06955162 &   0.99959703 &    89 &   0.15409613 &   0.99910052\\
   90 &   0.02627174 &   0.99980719 &    91 &   0.05201613 &   0.99982743 &    93 &   0.02655999 &   0.99976057\\
  100 &   0.05940698 &   0.99960737 &   101 &   0.02383782 &   0.99962639 &   105 &   0.09020480 &   0.99883707\\
  106 &   0.21004800 &   0.99933007 &   107 &   0.04443343 &   0.99971350 &   108 &   0.22005440 &   0.99814331\\
  109 &   0.00061440 &   0.99678018 &   110 &   0.05300864 &   0.93940588 &   111 &   0.00057600 &   0.99680101\\
  112 &   0.04628774 &   0.99969763 &   113 &   0.05432000 &   0.99955078 &   114 &   0.03080102 &   0.99678521\\
  148 &   0.01500288 &   0.99961681 &   149 &   0.00003840 &   0.99980076 &   150 &   0.02185856 &   0.99906731\\
  151 &   0.00975744 &   0.99923390 &   152 &   0.03208576 &   0.98612116 &   162 &   0.02767552 &   0.99025248\\
  166 &   0.02302374 &   0.99925389 &   167 &   0.04348672 &   0.99771274 &   168 &   0.02016000 &   0.99797157\\
  169 &   0.02720177 &   0.99795831 &   170 &   0.03968243 &   0.99648264 &   171 &   0.01991540 &   0.99450240\\
  174 &   0.44638592 &   0.73353986 &   180 &   0.02654400 &   0.99984933 &   181 &   0.01275392 &   0.99982848\\
  182 &   0.04051200 &   0.99963404 &   183 &   0.00333120 &   0.99966716 &   184 &   0.04056064 &   0.99957725\\
  185 &   0.00035840 &   0.99969758 &   186 &   0.00280320 &   0.99964336 &   187 &   0.02400000 &   0.99940374\\
  188 &   0.04047859 &   0.99950052 &   189 &   0.00615680 &   0.99978184 &   190 &   0.03057549 &   0.99960029\\
  191 &   0.00357504 &   0.99964138 &   192 &   0.01053888 &   0.99976005 &   193 &   0.03908733 &   0.99953959\\
  194 &   0.00004352 &   0.99967695 &   195 &   0.03335808 &   0.99973882 &   196 &   0.00033600 &   0.99983866\\
  197 &   0.00999808 &   0.99975767 &   198 &   0.03915146 &   0.99971150 &   199 &   0.00014400 &   0.99968692\\
  200 &   0.01800061 &   0.99930540 &   201 &   0.04000039 &   0.99923761 &   202 &   0.00109440 &   0.94475288\\
  203 &   0.00040320 &   0.98224294 &   204 &   0.00040320 &   0.98350837 &   205 &   0.02495764 &   0.99861626\\
  206 &   0.04869158 &   0.99660357 &   209 &   0.06209587 &   0.99816191 &   210 &   0.02496061 &   0.99949544\\
  211 &   0.02248128 &   0.99843373 &   213 &   0.02280640 &   0.99886374 &   214 &   0.03585600 &   0.99947434\\
  215 &   0.03319959 &   0.99870161 &   216 &   0.01588454 &   0.99971908 &   217 &   0.02455949 &   0.99934710\\
  218 &   0.02256310 &   0.99957031 &   219 &   0.01848448 &   0.99852887 &   220 &   0.03585434 &   0.99878318\\
  221 &   0.01704064 &   0.99891030 &   222 &   0.00279040 &   0.99960469 &   223 &   0.00438464 &   0.99974810\\
  224 &   0.00305152 &   0.99973111 &   225 &   0.03163648 &   0.98530041 &   226 &   0.00255424 &   0.99963640\\
  227 &   0.01775857 &   0.99912079 &   228 &   0.03187200 &   0.98588217 &   229 &   0.02007872 &   0.99853946\\
  231 &   0.01839987 &   0.99783360 &   232 &   0.02089562 &   0.93382328 &   233 &   0.00115840 &   0.99530680\\
  235 &   0.05586765 &   0.99853704 &   236 &   0.02767962 &   0.99926604 &   237 &   0.04774144 &   0.99683978\\
  238 &   0.07744878 &   0.99701717 &   239 &   0.00137600 &   0.99406434 &   240 &   0.00014400 &   0.99944023\\
  241 &   0.02630454 &   0.99948468 &   242 &   0.04112008 &   0.99837521 &   243 &   0.06197616 &   0.99894968\\
  244 &   0.03807885 &   0.99936549 &   245 &   0.02841963 &   0.99961186 &   246 &   0.04015885 &   0.99933769\\
  247 &   0.04867192 &   0.99846659 &   248 &   0.01520269 &   0.99926609 &   249 &   0.07598400 &   0.99815185\\
  250 &   0.01463946 &   0.99894841 &   251 &   0.03551923 &   0.99936410 &   252 &   0.07010816 &   0.99566558\\
  253 &   0.13527808 &   0.99512907 &   254 &   0.09176000 &   0.98946656 &   255 &   0.13588685 &   0.99096834\\
  256 &   0.19204966 &   0.99107519 &   258 &   0.25511360 &   0.94692244 &   259 &   0.00201600 &   0.99982368\\
  262 &   0.04304189 &   0.99884941 &   263 &   0.04536051 &   0.99711482 &   265 &   0.03627402 &   0.99870640\\
  266 &   0.00261504 &   0.99799184 &   267 &   0.02452928 &   0.99871793 &   268 &   0.02715034 &   0.99941643\\
  269 &   0.03500864 &   0.99656332 &   270 &   0.00237696 &   0.99926833 &   271 &   0.05063949 &   0.99905306\\
  272 &   0.01263424 &   0.99836319 &   273 &   0.00201920 &   0.99964485 &   274 &   0.04232384 &   0.99406479\\
  275 &   0.03899205 &   0.99970788 &   276 &   0.01303616 &   0.99969931 &   277 &   0.00238400 &   0.99956805\\
  278 &   0.00319360 &   0.99856571 &   279 &   0.01760128 &   0.99705974 &   280 &   0.04159532 &   0.99910513\\
  281 &   0.00544187 &   0.99524449 &   282 &   0.11486720 &   0.99604670 &   283 &   0.01346880 &   0.99663199\\
  284 &   0.13283064 &   0.98812420 &   285 &   0.10988800 &   0.99058653 &   286 &   0.26952000 &   0.97523832\\
  287 &   0.17388800 &   0.96681431 &   288 &   0.12025600 &   0.96621562 &   289 &   0.18153216 &   0.97157431\\
\hline
\end{tabular}\label{table2}
\end{table*}
\begin{table*}
\centering
\caption{Continuing table \ref{table2}.}
\begin{tabular}{p{1.15cm}<\centering |p{2.15cm}<\centering |p{2.15cm}<\centering || p{1.15cm}<\centering |p{2.15cm}<\centering |p{2.15cm}<\centering || p{1.15cm}<\centering |p{2.15cm}<\centering |p{2.15cm}<\centering}
\hline
SXS id & $e_{0}$ & $\text{FF}$ & SXS id & $e_{0}$ & $\text{FF}$ & SXS id & $e_{0}$ & $\text{FF}$\\
\hline
  290 &   0.25223910 &   0.97308490 &   291 &   0.00204800 &   0.95865500 &   292 &   0.26000000 &   0.96335086\\
  294 &   0.00744192 &   0.99977500 &   295 &   0.02647667 &   0.99981165 &   296 &   0.02122035 &   0.99975675\\
  297 &   0.01679539 &   0.99960447 &   298 &   0.04270400 &   0.99970251 &   299 &   0.02390400 &   0.99922898\\
  300 &   0.02569728 &   0.99805287 &   301 &   0.02656013 &   0.99964430 &   302 &   0.00977920 &   0.99970564\\
  303 &   0.02136000 &   0.99958484 &   304 &   0.07887058 &   0.99883377 &   305 &   0.00205120 &   0.99987321\\
  306 &   0.00017600 &   0.99719868 &   307 &   0.00568064 &   0.99898506 &   317 &   0.10803200 &   0.97590784\\
  318 &   0.01165824 &   0.99975644 &   319 &   0.00209574 &   0.99968403 &   320 &   0.13994860 &   0.99942937\\
  321 &   0.25980160 &   0.99928545 &   322 &   0.30415680 &   0.99799758 &   323 &   0.42212480 &   0.99661126\\
  324 &   0.56857178 &   0.95374415 &  1355 &   0.28001600 &   0.99678571 &  1356 &   0.35003200 &   0.99709325\\
 1357 &   0.43389120 &   0.99499395 &  1358 &   0.46010880 &   0.99423999 &  1359 &   0.43973760 &   0.99487680\\
 1360 &   0.52666240 &   0.99029515 &  1361 &   0.53544000 &   0.99066395 &  1362 &   0.59000000 &   0.98289242\\
 1363 &   0.59000000 &   0.94543609 &  1364 &   0.26428480 &   0.99858436 &  1365 &   0.32008621 &   0.99731353\\
 1367 &   0.39434637 &   0.97010512 &  1368 &   0.45011761 &   0.99801954 &  1369 &   0.58760000 &   0.97968781\\
 1370 &   0.58371840 &   0.93113201 &  1371 &   0.28035200 &   0.99668417 &  1372 &   0.45632640 &   0.99914158\\
 1373 &   0.48023360 &   0.99043551 &  1374 &   0.58482240 &   0.87252240 &  1426 &   0.00235162 &   0.93906894\\
 1429 &   0.04240051 &   0.99936212 &  1430 &   0.02640000 &   0.99586796 &  1431 &   0.01600128 &   0.99884619\\
 1432 &   0.01227023 &   0.85881671 &  1436 &   0.04896000 &   0.99869367 &  1438 &   0.04838285 &   0.99819549\\
 1439 &   0.00821760 &   0.83222335 &  1441 &   0.03095744 &   0.90245143 &  1443 &   0.01119680 &   0.97523844\\
 1444 &   0.00319680 &   0.99774492 &  1445 &   0.01663762 &   0.97814507 &  1448 &   0.02841562 &   0.99861477\\
 1450 &   0.01409600 &   0.99231581 &  1451 &   0.03911693 &   0.98826061 &  1453 &   0.26915200 &   0.92573267\\
 1455 &   0.03023808 &   0.99059889 &  1457 &   0.45003200 &   0.95301673 &  1458 &   0.00928640 &   0.97538328\\
 1460 &   0.00091712 &   0.98562490 &  1461 &   0.00241190 &   0.99782163 &  1463 &   0.08521408 &   0.91667173\\
 1464 &   0.02143424 &   0.98786709 &  1465 &   0.02840000 &   0.99682826 &  1466 &   0.17256832 &   0.98473899\\
 1467 &   0.00008128 &   0.98519339 &  1468 &   0.04248000 &   0.99513143 &  1474 &   0.01080000 &   0.99851301\\
 1476 &   0.02440000 &   0.99769347 &  1478 &   0.23993600 &   0.95487160 &  1479 &   0.02335744 &   0.99833449\\
 1480 &   0.00081216 &   0.99850227 &  1482 &   0.04080320 &   0.99571523 &  1483 &   0.04974400 &   0.96844915\\
 1484 &   0.04053128 &   0.99357910 &  1485 &   0.02183898 &   0.99478509 &  1486 &   0.04796544 &   0.98702723\\
 1487 &   0.00351360 &   0.99642889 &  1488 &   0.00232512 &   0.99613647 &  1489 &   0.08024000 &   0.99394981\\
 1490 &   0.01336832 &   0.99095196 &  1491 &   0.04731072 &   0.99842423 &  1492 &   0.00152640 &   0.99911536\\
 1493 &   0.02866880 &   0.99458259 &  1494 &   0.00117440 &   0.99849727 &  1495 &   0.00129280 &   0.95701687\\
 1496 &   0.02040064 &   0.99756563 &  1497 &   0.02160192 &   0.94379466 &  1498 &   0.00438259 &   0.99949146\\
 1499 &   0.02764646 &   0.99933705 &  1500 &   0.00038720 &   0.99925089 &  1501 &   0.03176179 &   0.99765457\\
 1502 &   0.03540800 &   0.99807781 &  1503 &   0.00016000 &   0.99757773 &  1504 &   0.09433600 &   0.99868777\\
 1505 &   0.04240026 &   0.99724187 &  1506 &   0.00822848 &   0.99900658 &  1507 &   0.00440000 &   0.99858852\\
 1508 &   0.05049549 &   0.99915065 &  1509 &   0.02385050 &   0.99962697 &       &              &\\
\hline
\end{tabular}\label{table3}
\end{table*}

As we analyzed in the above section, the SEOBNRE model fits the NR waveforms quite well for spin-aligned BBH systems. For systems with mild initial eccentricity $e_0$ and mild spin hang-up effect, the fitting factor can be as good as 99\%. For gravitational wave detection practice, fitting factor $>99$\% is likely to be enough \cite{PhysRevD.95.044028}. It is interesting to improve the SEOBNRE model to expand the validity domain of the eccentricity and the spin hang-up effect. This problem is out of the scope of this paper and we leave it to future work.

Due to the gauge problem and the strong general relativity effect \cite{Loutrel:2018ydu} involved in the initial data of NR simulations, it is not definite in determining the initial eccentricity with NR technique \cite{Pfeiffer_2007,PhysRevD.77.044037,PhysRevD.79.124040,PhysRevD.82.124016,PhysRevD.83.024012,PhysRevD.83.104034,PhysRevD.85.124051}. This poorly determined quantity will introduce an uncertainty when we use it to construct the gravitational waveform template bank. Based on our accurate waveform model which can account for the initial eccentricity, we use the SEOBNRE model to calibrate the NR waveforms on the initial eccentricity. More importantly, we can set a quite low common reference frequency, say $Mf_0=0.002$, which corresponds to a large separation of the two inspiralling black holes. This kind of setting makes sure that the adiabatic approximation is valid and the concept eccentricity is meaningful.

The initial eccentricity at the reference frequency $Mf_0=0.002$ and the fitting factor between NR waveforms and the SEOBNRE model for spin-aligned BBHs in the SXS simulation bank \cite{SXSBBH} are listed in Tables \ref{table2}-\ref{table3}. Here we have investigated 278 waveforms all together. These waveforms include BBHs with mass ratio from 1 to 10 and with a largest aligned spin of 0.995. We obtain fitting factors larger than 99\%
for most waveforms.

\section{Summary}
Although it is not clear yet whether there are BBH sources for the advanced LIGO/Virgo moving along eccentric orbit \cite{Hoang_2019}, there exist some formation channels that could produce eccentric BBHs \cite{strader2012two,samsing2014formation,vanlandingham2016role,antonini2016black,Zhang:2019puc}. In the near future there will definitely be many BBH sources for space-based detectors that admit significant orbit eccentricities \cite{hils1995gradual,PhysRevD.50.6297,Armitage_2005}. Recently, more and more researchers care about compact object binary systems with eccentric orbit regarding to gravitational wave detection \cite{PhysRevD.80.084001,PhysRevD.86.104027,PhysRevD.87.127501,PhysRevD.90.084016,PhysRevD.90.104010,PhysRevD.91.063004,PhysRevD.92.044034,PhysRevD.96.084046,PhysRevD.100.064006,Loutrel:2019kky,Loutrel:2016cdw,Loutrel:2017fgu,Loutrel:2018ssg,Moore:2018kvz,Loutrel:2018ydu,Gondan:2018khr,Gondan:2017wzd,Hoang:2017fvh,Gondan:2017hbp}.

In order to make matched filtering data analysis technique work, many gravitational waveform models have been proposed. Based on the assumption of a low eccentricity, Ref.~\cite{PhysRevD.80.084001} extended low order PN waveform model in frequency domain to include the orbital eccentricity. They named the corresponding model the post-circular (PC) model. Ref.~\cite{PhysRevD.90.084016} phenomenologically extended the PC model to the enhanced post-circular (EPC) model which recovers the TaylorF2 model when the eccentricity vanishes. The overall PN order of the EPC model is 3.5. Some NR simulations have been carried out in the past for eccentric BBH systems \cite{PhysRevD.77.081502,PhysRevD.78.064069,PhysRevD.82.024033,PhysRevD.87.043004,PhysRevD.88.064051,PhysRevD.100.044016,PhysRevD.100.064003,Ramos-Buades:2019uvh}. Combined with the NR results, the x-model was proposed in Ref.~\cite{PhysRevD.82.024033}. The x-model is a low order post-Newtonian (PN) model. The x-model was recently developed to the advanced x-model (ax-model), which includes the higher PN order terms for quasi-circular part, to cover inspiral, merger and ringdown phases \cite{PhysRevD.95.024038}. All these models use the eccentricity as an explicit quantity to describe BBH's motion which means that they all take the adiabatic approximation.

In contrast, we treat the eccentric BBH systems within the EOBNR framework which makes us avoid the adiabatic approximation. We constructed the SEOBNRE model in \cite{PhysRevD.96.044028} which is consistent to SEOBNR for quasi-circular binary cases. When quasi-circular BBH systems are considered, for the mass ratio ranging in [1,10] and the aligned spin ranging in [0,0.8], the consistency between the SEOBNRE model and SEOBNR models is better than 99.98\%.

In this paper we validate our SEOBNRE model to the NR simulation results \cite{SXSBBH}. For generic eccentric spin-aligned BBH systems, there are 4 parameters, including the initial eccentricity $e_0$, the mass ratio $q$, the effective spin $\chi_{\rm eff}$ and the anti-symmetric spin $\chi_A$. According the comparison result between the SEOBNRE model and the NR waveforms, we find the limitations of the SEOBNRE model come from the initial eccentricity $e_0$ and the spin hang-up effect. The limitation on the initial eccentricity is $e_0<0.55$ at the reference frequency $Mf_0=0.002$. The spin hang-up effect can be described by a combined parameter which we call the spin hang-up parameter. The spin hang-up parameter is defined in Eq.~(\ref{equation4}). The limitation on this spin hang-up parameter is $\chi_{\rm up}<0.4$.

It is quite hard to determine the initial eccentricity in NR. Many clever methods are developed to reduce and/or determine the initial eccentricity in NR \cite{Pfeiffer_2007,PhysRevD.77.044037,PhysRevD.79.124040,PhysRevD.82.124016,PhysRevD.83.024012,PhysRevD.83.104034,PhysRevD.85.124051,Ramos-Buades:2019uvh}. Still, some NR simulations, such as SXS:BBH:0071, SXS:BBH:1362, SXS:BBH:1363, SXS:BBH:0148, SXS:BBH:0151, SXS:BBH:0170, SXS:BBH:0171 and more, can not determine the initial eccentricity properly. As an extra usage, our SEOBNRE model can be used to calibrate the initial eccentricity of the NR simulation results. We hope this kind of calibration will help to improve the parameter estimation in the gravitational wave data analysis.

Currently we did not introduce extra adjustable parameters when we construct SEOBNRE models, compared to the corresponding SEOBNR models. In future we could apply the technique implemented in Ref.~\cite{PhysRevD.95.044028} to introduce adjustable parameters, and determine these parameters based on the calibration of the waveform to NR results. In this mean we can determine the adjustable parameters for circular part and eccentric part altogether. Hopefully this may improve our SEOBNRE model to alleviate or even remove the limitation on the initial eccentricity and the spin hang-up effect.

\section*{Acknowledgments}
This work was supported by the NSFC (No.~11690023, No.~11622546 and No. 11975027). Z. Cao was supported by ``the Fundamental Research Funds for the Central Universities", ``the Interdiscipline Research Funds of Beijing Normal University" and the Strategic Priority Research Program of the Chinese Academy of Sciences, grant No. XDB23040100. L. Shao was supported by the Young Elite Scientists Sponsorship Program by the China Association for Science and Technology (2018QNRC001).

\begin{widetext}
\appendix
\section{Post-Newtonian expression of the orbit decay for quasi-circular binary} \label{App}
With the relation between the energy and the semimajor axis, $E=-{\eta M^2}/{2a}$, and the relation between the variable $x$ (defined in Ref.~\cite{PhysRevD.96.044028}) and the semimajor axis, $x={M}/{a}$, we
have
\begin{align}
E=-\frac{M}{2}x.
\end{align}
Using the above equation and Eq.~(C1) in Ref.~\cite{PhysRevD.96.044028}, we
can obtain the orbit decay rate as following,
\begin{align}
\frac{da}{dt}=&-\frac{64}{5}\eta^2 x^3[1-\frac{1247+980\eta}{336}x+4\pi x^{3/2}+(-\frac{44711}{9072}+\frac{9271\eta}{504}+\frac{65\eta^2}{18})x^2
-(\frac{8191}{672}+\frac{583\eta}{24})\pi x^{5/2}\nonumber\\
&+(\frac{6643739519}{69854400}-\frac{1712\gamma_E}{105}+\frac{16 \pi ^2}{3}-\frac{134543 \eta }{7776}+\frac{41 \pi ^2 \eta }{48}-\frac{94403 \eta ^2}{3024}-\frac{775 \eta ^3}{324}-\frac{856 \log(16)}{105}-\frac{856 \log(x)}{105})x^3\nonumber\\
&+(-\frac{16285}{504}+\frac{214745\eta }{1728}+\frac{27755\eta ^2}{432})\pi x^{7/2}\nonumber\\
&+(-\frac{23971119313}{93139200}+\frac{856 \gamma_E}{35}-8 \pi ^2-\frac{59292668653 \eta }{838252800}+5 a_0 \eta +\frac{856 \gamma \eta }{315}+\frac{31495 \pi ^2 \eta }{8064}-\frac{54732199 \eta ^2}{93312}+\frac{3157 \pi ^2 \eta ^2}{144}\nonumber\\
&+\frac{18929389 \eta ^3}{435456}+\frac{97 \eta ^4}{3888}+\frac{428 \log(16)}{35}+\frac{428}{315} \eta  \log(16)+\frac{428 \log(x)}{35}+\frac{47468}{315} \eta  \log(x))x^4\nonumber\\
&+(-\frac{80213 }{768}+\frac{51438847  \eta }{48384}-\frac{205 \pi ^2 \eta }{6}-\frac{42745411  \eta ^2}{145152}-\frac{4199 \eta ^3}{576})\pi x^{9/2}\nonumber\\
&+(-\frac{121423329103}{82790400}+\frac{5778 \gamma_E}{35}-54 \pi ^2+\frac{4820443583363 \eta }{1257379200}-\frac{3715 a_0 \eta }{336}+6 a_1 \eta -\frac{4066 \gamma_E \eta }{35}-\frac{31869941 \pi ^2 \eta }{435456}\nonumber\\
&-\frac{2006716046219 \eta ^2}{3353011200}-\frac{55 a_0 \eta ^2}{4}+\frac{214 \gamma_E \eta ^2}{105}+\frac{406321 \pi ^2 \eta ^2}{48384}+\frac{2683003625 \eta ^3}{3359232}-\frac{100819 \pi ^2 \eta ^3}{3456}-\frac{192478799 \eta ^4}{5225472}\nonumber\\
&+\frac{33925 \eta ^5}{186624}+\frac{2889 \log(16)}{35}-\frac{2033}{35} \eta  \log(16)+\frac{107}{105} \eta ^2 \log(16)+\frac{2889 \log(x)}{35}-\frac{391669}{315} \eta  \log(x)\nonumber\\
&-\frac{122981}{105} \eta ^2 \log(x))x^5\nonumber\\
&+(-\frac{623565 }{1792}-\frac{235274549 \eta }{241920}+20 a_0  \eta +\frac{852595 \pi ^2 \eta }{16128}-\frac{187219705 \eta ^2}{32256}+\frac{12915 \pi ^2 \eta ^2}{64}+\frac{503913815 \eta ^3}{870912}\nonumber\\
&-\frac{24065 \eta ^4}{3456}+\frac{1792}{3} \eta  \log(x)) \pi x^{11/2}\nonumber\\
&+(-\frac{1216355221}{206976}+\frac{4815 \gamma_E}{7}-225 \pi ^2+\frac{45811843687349 \eta }{1149603840}+\frac{170515 a_0 \eta }{18144}-\frac{743 a_1 \eta }{56}+7 a_2 \eta +a_3 \eta \nonumber\\
&-\frac{737123 \gamma_E \eta }{189}-\frac{84643435883 \pi ^2 \eta }{670602240}+\frac{8774}{63} \gamma_E \pi ^2 \eta -\frac{410 \pi ^4 \eta }{9}-\frac{37516325949517 \eta ^2}{603542016}+\frac{68305 a_0 \eta ^2}{2016}-\frac{33 a_1 \eta ^2}{2}\nonumber\\
&+\frac{6634 \gamma_E \eta ^2}{63}+\frac{23084972185 \pi ^2 \eta ^2}{5225472}-\frac{92455 \pi ^4 \eta ^2}{1152}+\frac{6069288163291 \eta ^3}{2586608640}+\frac{295 a_0 \eta ^3}{18}+\frac{107 \gamma_E \eta ^3}{243}-\frac{114930545 \pi ^2 \eta ^3}{1741824}\nonumber\\
&-\frac{145089945295 \eta ^4}{282175488}+\frac{141655 \pi ^2 \eta ^4}{7776}+\frac{6942085 \eta ^5}{497664}+\frac{196175 \eta ^6}{3359232}+\frac{4815 \log(16)}{14}-\frac{737123}{378} \eta  \log(16)\nonumber\\
&+\frac{4387}{63} \pi ^2 \eta  \log(16)+\frac{3317}{63} \eta ^2 \log(16)+\frac{107}{486} \eta ^3 \log(16)+\frac{4815 \log(x)}{14}+\frac{13185899 \eta  \log(x)}{59535}+7 a_3 \eta  \log(x)\nonumber\\
&+\frac{4387}{63} \pi ^2 \eta  \log(x)+\frac{963937}{189} \eta ^2 \log(x)+\frac{6279367 \eta ^3 \log(x)}{2430})x^6\nonumber\\
&+(-2\chi_S-\frac{3}{4}\sqrt{1-4\eta}\chi_A)x^{3/2}\nonumber\\
&+[(-\frac{9}{4}+\frac{136}{9}\eta)\chi_S+(-\frac{23}{16}
+\frac{157}{36}\eta)\sqrt{1-4\eta}\chi_A]x^{5/2}\nonumber\\
&+(-8\pi\chi_S-\frac{17}{6}\pi\sqrt{1-4\eta}\chi_A)x^3\nonumber\\
&+((\frac{476645}{13608}+\frac{3086}{189}\eta-\frac{1405}{27}\eta^2)\chi_S
+(\frac{180955}{27216}+\frac{625}{378}\eta-\frac{1117}{108}\eta^2)\sqrt{1-4\eta}\chi_A)x^{7/2}],\label{equation3}
\end{align}
with $a_0=153.8803,a_2=-55.13,a_2=588,a_3=-1144$.
\end{widetext}
\bibliography{refs}

\end{document}